\documentstyle[preprint,aps,tighten]{revtex}
\begin{document}

\title{``TelePOVM'' -- More Faces of Teleportation}

\author{Tal Mor\thanks{Physics Department, Technion, Haifa 32000, Israel}}

\date{\today}

\maketitle

\begin{abstract}

We show two new aspects of teleportation based on 
generalized measurements:
I. We present an alternative approach to teleportation 
based on ``generating $\rho$-ensembles at spacetime separation'';
This approach provides a better understanding of teleportation.
II. We define the concept of {\em conclusive teleportation} and
we show how to teleport a quantum state using 
any pure entangled state; 
we show how to use the conclusive teleportation as a criterion 
for non-locality.

\end{abstract}

\section{INTRODUCTION}

Suppose that Alice has an unknown quantum state 
(say in 2-dimensional Hilbert space, ${\cal H}_2$)
\begin{equation} {a \choose b} =  a |\uparrow \rangle
+ b |\downarrow \rangle  \end{equation}
which she wishes to transmit to Bob.
Surprisingly,
the minimal resources required 
in order to perform this are 
one EPR singlet pair and two classical 
bits~\cite{BBCJPW}: 
Alice and Bob distribute an EPR pair so that each 
has one particle; Alice performs a Bell measurement on her particle
and the unknown quantum state; She sends the two-bits result to Bob
who uses it to recover the unknown quantum state from his member of the EPR
pair by simple rotation.
This process, which received the name {\em teleportation}, is mysterious 
because a spin state of a spin-half
particle is described by a point on a unit sphere , hence by two real numbers;
Yet, an EPR state (which is independent of the spin-state) and two
classical bits are enough for such a transmission.

The alternative approach presented in section 2 clarifies the mystery. 
It also explains 
where the surprising number -- $2$ bits, comes from. 
Our approach is based on a recent work 
on classification of quantum ensembles having a given density matrices
\cite{HJW}.
An extremely important result of this 
paper generalizes the EPR paradox and states the following: 
suppose Alice and Bob share an entangled state and the reduced density matrix
in Bob's hands is $\rho$. 
Let $\{\rho\}$ (called a $\rho$-ensemble) 
be a set of pure states $\{\psi_j\}$ 
and probabilities
${p_j}$, such that
$\ \sum_j p_j \psi_j = \rho$.
Alice can generate any $\rho$-ensemble in Bob's hands by performing operations
far away. 
(This is referred to as 
``generating $\rho$-ensembles at spacetime separation'' in that paper.)
The meaning of this result is the following:
Alice performs a specific ganeralized measurement (POVM) 
at her site (in general, using an ancila), 
which yields one of $m$ possible results with some probabilities
$p_i$, ($\sum_{i=1}^m p_i = 1$);
Doing this, 
Alice creates the same situation as if she sent Bob one of $m$ possible states
in $\{\psi_j\}$ with the probabilities as above.
We show that the teleportation is a special case of generating 
$\rho$-ensembles at a distance, in which Alice's measurement is done
using an ancila in some (unknown) initial state ${a \choose b}$, 
and the state of the ancila can 
{\em always} be recovered at Bob's site from any possible state in $\{\psi_j\}$.
This explains teleportation in terms of measurement theory 
and reduces its mystery.

The next natural step is to use this approach to generalize the concept of 
teleportation, by removing the demand that the transmitted state
can always be recovered.
In section 3 we define the concept of {\em conclusive teleportation}.
The term ``conclusive'' is taken from quantum information theory, where
one asks the following question: 
what is the optimal mutual information which can be extracted from 
two quantum states?
For two orthogonal states it is $1$ bit.
For two non-orthogonal states the optimal information is less than $1$ bit,
and it is derived by a standard measurement;
the identification of the state is non-perfect and there is
some probability
of mistake.
Alternatively, one can obtain a definite (correct) answer \underline{sometimes}
for the price of knowing nothing (or very little) 
in other occasions~\cite{Per93}.
Adapting this term to teleportation we present the conclusive teleportation 
in which the teleportation process is \underline{sometimes} successful, and the 
sender knows if it is or not.
When Alice and Bob use 
entangled pure state which is not fully entangled
the conclusive teleportation scheme allows them to teleport a quantum state
with fidelity one.
This is done for the price of occasional failures, and 
the sender knows whether it is successful or it is not.
For most purposes (e.g., for quantum cryptography~\cite{BB84,EPR,Ben92,BHM}), 
one would prefer performing 
this conclusive teleportation rather than performing the Bell
measurement (as in the standard 
teleportation scheme) and teleporting the unknown state with fidelity smaller 
than one  
(the received state differs from the original state)
as previously suggested by Gisin~\cite{Gisin}. 
 
\section{Telepovm}

The teleportation process can be 
understood using the concept of 
generalized measurements~\cite{JP-DL,Helstrom,Per93} 
performed on the partial Hilbert space, ${\cal H}_2$, which is in Alice's hand.
Suppose that Alice and Bob share 
any pure state in any dimension, such that the reduced density
matrix in Bob's hands is $\rho$. 
Any measurement at Alice side, performed on her part of the entangled state
creates a specific $\rho$-ensemble in Bob's hands.
This is proved in~\cite{HJW} (henceforth, HJW). 
Since all $\rho$-ensembles are indistinguishable (recall that a quantum system
is fully described by its density matrices) the concept of $\rho$-ensemble
requires further clarification:
Two quantum systems having the same density matrices {\em can be 
distinguished} if there exist an additional information somewhere.
For example, in the BB84 cryptographic scheme~\cite{BB84}
Bob receives the same density 
matrix $\rho$ whether Alice uses the $x$-basis or the $z$-basis, but he receives
different $\rho$-ensembles. He cannot distinguish between the two ensembles
and between the states in each particular occasion, 
unless he receives more information from Alice. 
When receiving additional information (the basis) he is told which 
$\rho$-ensemble he has, and can find which state. 
In the same sense, the EPR-scheme\cite{EPR}, provides an example
of the HJW meaning of $\rho$-ensembles:
when Alice chooses to measure her member of the singlet
state in the $x$-basis or in the $z$-basis, she forces a different
$\rho$-ensemble in Bob's hands. He can distinguish the two states 
to find Alice's bit 
after receiving additional information from Alice who tells him the basis
(hence -- the $\rho$-ensemble).
Alice's choice of measurement determines the $\rho$-ensemble, and her result,
in each occasion, tells her which of the states
is in Bob's hands.

This trivial example was generalized in HJW 
to include also the case where the standard, 
projection measurement is replaced by a generalized 
measurement~\cite{JP-DL,Helstrom,Per93} 
(also called ``positive operature value measured'' -- POVM),
so the number of results can be larger than the dimension
of the Hilbert space in Alice's site or in Bob's site.
Unlike the previous case, Alice's particle is not in a well defined
pure state anymore (e.g., since it is measured together with an ancilla --
see ahead).
Yet, Bob's particle is in a well-defined state. Alice's choice of the POVM  
forces a $\rho$-ensemble. The result she obtains tells her
which state from the ensemble is in Bob's hands. 
This is a very interesting result of~\cite{HJW} and we soon show 
that teleportation provides  
a fascinating usage of it.
On the other hand, the calculation presented in the teleportation 
paper~\cite{BBCJPW} 
provides another proof that indeed a $\rho$-ensemble is created, thus
-- a verification of the result of HJW.

Let Alice and Bob share an EPR pair.
Consider the following POVM:
\begin{equation}  A_1 = \frac{1}{2}  \left( \begin{array}{cc} 
                          c^2 & cs \\ cs & s^2 
\end{array} \right)  \ ;
     A_2 = \frac{1}{2}  \left( \begin{array}{cc} 
                          c^2 & -cs \\ -cs & s^2 
\end{array} \right)  \ ;
\end{equation}
\begin{equation}  A_3 = \frac{1}{2}  \left( \begin{array}{cc} 
                          s^2 & cs \\ cs & c^2 
\end{array} \right)  \ ;
     A_4 = \frac{1}{2}  \left( \begin{array}{cc} 
                          s^2 & -cs \\ -cs & c^2 
\end{array} \right)  \ ,
\end{equation}
with $s \equiv \sin\theta$ and $c\equiv \cos \theta$.
These matrices have positive eigenvalues and
sum up to the unit matrix therefore form a POVM.
Following the arguments of HJW, 
applying such a POVM to one member of two particles
in an EPR state is equivalent to a choice of a specific
$\rho$-ensemble combined of four possible states; 
when the result of the POVM is $A_i$, the other member
is projected onto a state orthogonal to $A_i$,
i.e., it will be in one of the states 
$\psi_1 = {s \choose -c}\ $;
$\psi_2 = {s \choose c}\ $;
$\psi_3 = {c \choose -s}\ $, and 
$\psi_4 = {c \choose s}$ respectively, and Alice will know in which
of them. Alice can tell Bob the result of her measurement, which is actually
telling him which of them he got. This is done by sending 
him two classical bits.
The uniqueness of that POVM is that the four resulted states are connected
in a special way -- which is {\em independent} of the angle $\theta$:
Bob can re-derive one of the states, say ${c \choose s}$,
by performing a simple operation of rotation around
one of the main axes (or do nothing), 
according to the two classical bits he is being told.

Every POVM can be performed in the lab by performing a standard measurement
on the system $\rho_{sys}$, plus an ancilla \cite{IvPe,Per93}
(this is a property of the POVM
so it is true independently
of the state of the system).
One way to perform the POVM just described is to take an ancilla
in ${\cal H}_2$, in an initial state ${c \choose s}$,
and perform the Bell measurement on the ancilla and the system.
We prove this using a cumbersome technique which is 
explained in~\cite{Per93} (Chapter 9, 
subsections 9.5 and 9.6 which discusses generalized measurements
and Neumark's theorem).
The first operator, $A_1$, 
results from the measurement of the projection operator
$P_1 = | \Phi_+ \rangle \langle \Phi_+ |$ in the space of Alice's particle
plus the ancilla. Its terms are  
\begin{equation} (A_1)_{mn} = \sum_{rs} (P_1)_{mr,ns} (\rho_{aux})_{sr} 
\ , \end{equation}
where $\rho_{aux}$ is the state of the ancilla, the $mn$ are the
indices of the particle
and the $sr$ are indices of the ancilla.
The $m=0,\ n=0$ case corresponds to multiplying the 
upper left block of $P_1$ by the density matrix of the ancilla, and 
tracing the obtained matrix yielding:
\begin{equation} \sum_{sr} 
\left(\begin{array}{cc} \frac{1}{2} & 0 \\ 0 & 0
\end{array}\right)_{rs}
\left(\begin{array}{cc} c^2 & cs \\ cs & s^2
\end{array}\right)_{sr} = \frac{1}{2} c^2 \ .
\end{equation}
In the same way we calculated the other elements of that operator,
and the other three operators, and we verified that 
the Bell measurement corresponds to the desired POVM.
As previously said (and showed in the teleportation paper), 
the Bell measurement creates  
one of four states in Bob's hands
and Bob can re-derive the original state of the ancilla by rotation, completing 
the ``telepovm'' process. 

The crucial point here is that Bob reproduce the original state
using one of four operations where neither of these operations depends on
the state of the ancilla (rotations around the principle axis).
Hence Alice needs only tell him the two bits result of her measurement,
to choose one of (previously agreed) four possible rotations.
This can be done even if Alice doesn't know the state of the ancilla, 
say ${c \choose s}$, and this is exactly the process of teleportation
of an unknown state.
No matter what the angle $\theta$ is, 
Alice's measurement generates the appropriate $\rho$-ensemble at Bob's site.
The process can be easily generalized to include 
ancilla in any pure state ${a \choose b}$ with complex numbers satisfying
$|a|^2 + |b|^2 =1$ (and will teleport also a density matrix or a particle in
an entangled state). We considered only the case of a spin state in the 
$xz$ plane for simplicity. It can also easily be generalized to fully
entangled states in higher ($N^2$) dimensions discussed in \cite{BBCJPW}.

To see one application of this, let us view a different scenario:
Suppose that Alice have in mind a set 
of states and their probabilities $\{\rho\}$
(as in the BB84, or many other  
quantum cryptography scheme).
If Alice doesn't care which of the states is sent in each experiment, but only
that it belongs to that set, i.e., to that $\rho$-ensemble, she does not 
need to send the states. Instead of sending Bob the states she sends him
a member of some entangled state such that the reduced density matrix in Bob's
hands is $\rho$ (she can ask someone else, even Eve, to create that state).
Than she applies a specific POVM which creates the desired ensemble in Bob's
hands. 
The well known example is the EPR scheme~\cite{EPR}.
We present another example:  when the state
\begin{equation} \Psi = \alpha | \uparrow_x \uparrow _x\rangle + 
              \beta | \downarrow_x \downarrow_x \rangle \ , \end{equation}
is measured by Alice (a standard measurement) in the $z$ direction,
the following $\rho$-ensemble is generated in Bob's hands:
\begin{equation} { \frac{\alpha+\beta}{\sqrt 2} \choose 
                                \frac{\alpha-\beta}{\sqrt 2} } 
\quad ; \quad
     { \frac{\alpha-\beta}{\sqrt 2} \choose 
                                \frac{\alpha+\beta}{\sqrt 2} } \ .
\end{equation}
This can be used to produce 
the B92 \cite{Ben92} scheme for quantum cryptography,
in the same way that the EPR scheme produces the BB84~\cite{BB84} scheme.

\section{Teleportation with any entangled state}

The fully entangled state can be used to perform a conclusive teleportation:
Let Alice perform a measurement which distinguishes the singlet state from 
the other three (triplet) states. Instead of sending $2$ bits she sends Bob
only one bit telling him whether she received a singlet state or not.
In a $\frac{1}{4} $ of the occasions indeed she receives this result,
hence performs a successful teleportation.
This process does not make much sense for this case (however, it saves one
classical bit), but makes a lot of 
sense when Alice and Bob share a pure entangled
state which is not fully entangled.

Let Alice and Bob share the state
\begin{equation} \Psi_{23} = \alpha | \uparrow \uparrow \rangle + 
              \beta | \downarrow \downarrow \rangle \  \end{equation}
(any pure state can be written in that form called the Schmidt decomposition),
which they use to teleport a quantum state ${a \choose b}_1$.
Following the method of \cite{BBCJPW}, the state of the three particles 
is written 
using the Bell states (defined in \cite{BMR}) as:
\begin{equation} \Psi_{123} = {a \choose b}_1 \Psi_{23} \\ =
 \Phi^+_{12} {\alpha a \choose \beta b}_3 +
 \Phi^-_{12} {\alpha a \choose - \beta b}_3 +
 \Psi^+_{12} {\alpha b \choose \beta a}_3 +
 \Psi^-_{12} {\alpha b \choose-  \beta a}_3 \ . \end{equation}
A Bell measurement still creates the same POVM as before.
But, unlike the case of using a fully entangled state,
the states created in Bob's hands depend also on $\alpha$ and $\beta$,
and not only on the state of the ancilla. Therefore, the teleported 
state cannot be reproduced by a simple rotation around the principle axis. 
Bob can still reproduce the original state if Alice tells him how to perform
the rotations, but this cannot be used to teleport an unknown quantum state.
It is also possible to find a POVM that 
reproduce the four desired states (those which can be 
rotated around the principle axis as before). 
However, the POVM is not performed by a Bell measurement and 
will depend on the state of
the ancilla.
For a given state (known to Alice), this is fine, but  
this is not a teleportation of an unknown state.

We present a different measurement which generates the desired states in Bob's 
hands. The price we pay for the perfect state obtained, is that the process
cannot be done with 100\% probability of success, therefore it is a conclusive
teleportation.
To explain how it works, 
let us return to the case of fully entangled state and
separate the Bell measurement into two measurements (one follows the other):
\begin{itemize} 
\item A measurement which checks whether the state is in the subspace
spanned by $ | \uparrow \uparrow \rangle $ and 
$ | \downarrow \downarrow \rangle $, or in the subspace spanned by 
$ | \uparrow \downarrow \rangle $ and 
$ | \downarrow \uparrow \rangle $. 
\item A measurement in the appropriate subspace (according to the result of the 
previous step), which projects the state on one of the two possible
Bell states in that subspace.
\end{itemize}
When $\Psi_{23}$ 
is not fully entangled we still repeat
the first step of that two-steps process.
The state of the three particles can also be written as
\begin{eqnarray} \Psi_{123} = 
    \frac{1}{2} [\alpha | \uparrow   \uparrow   \rangle 
            + \beta  | \downarrow \downarrow \rangle ] {a \choose b}_3
+   \frac{1}{2} [\alpha | \uparrow   \uparrow   \rangle 
            - \beta  | \downarrow \downarrow \rangle ] {a \choose -b}_3
\nonumber \\
+   \frac{1}{2} [\alpha | \uparrow   \downarrow \rangle 
            + \beta  | \downarrow \uparrow   \rangle ] {b \choose a}_3
+   \frac{1}{2} [\alpha | \uparrow   \downarrow \rangle 
            - \beta  | \downarrow \uparrow   \rangle ] {b \choose -a}_3
 \ ,\end{eqnarray}
and the first step projects it on either the first two possibilities or
the last two with equal probabilities.
Let us use either the basis
\begin{equation} | \uparrow \uparrow \rangle = {1 \choose 0}_{12} \quad  {\rm and}
\quad  | \downarrow \downarrow \rangle = {0 \choose 1}_{12} \end{equation}
or 
\begin{equation} 
| \uparrow \downarrow \rangle = {1 \choose 0}_{12}
\quad  {\rm and}
\quad  
| \downarrow \uparrow \rangle = {0 \choose 1}_{12} 
\ , \end{equation}
depending on the result of the first step.
We replace the second step, which was a projection on one of two
Bell states in one of the two-dimensional subspaces, 
by a POVM which can conclusively distinguish
the two possible states, 
${\alpha \choose \beta}_{12} $ and $ {\alpha \choose -\beta}_{12}$, 
where the basis depends on the result of the first step.

Such a process is well known \cite{Per93,EHPP} due to analysis of the 
two-states scheme \cite{Ben92}.
Using the optimal process, a conclusive result is obtained with probability 
$1 - (|\alpha|^2 - |\beta|^2)$, hence this is the probability
of a successful teleportation. 
Alice tells Bob whether she succeeded in teleporting the state by sending him
one bit.
In addition to this bit, 
Alice still has to send Bob the two bits for distinguishing the
four possible states (to perform the right rotation).
Alternatively, she can send him only one bit telling him whether he received
the state or not (as we explained for the case of fully entangled state)
loosing $\frac{3}{4}$ of the successful teleportations.
A POVM allows to get the optimal deterministic information
from two non-orthogonal states, although, on average, 
it yields less mutual information 
than the optimal projection measurement. In the same sense, on average,  
the conclusive teleportation
does not yield the optimal average fidelity, but when it is successful -- 
the fidelity is one.

The conclusive teleportation process proves that any pure-entangled state
is non-local (as an alternative to Bell-inequality violation):
it can be used for teleportation with fidelity one.
This cannot be done classically. The use of teleportation as a non-locality 
test was suggested by Gisin\cite{Gisin}. 
However, he discusses the original teleportation
scheme, which cannot yield fidelity one unless fully entangled states are used.
While his criterion recognizes non-locality of some mixed states better than
the Bell's inequalities criteria, it is inferior to Bell's criteria in other
cases: for example,
it cannot notice that any pure entangled state is non-local.

It is easy to think about a scenario where this can be used:
Let Alice try to teleport to Bob a quantum state known to another person, Carl.
She is allowed to 
inform Bob whether the 
teleportation is successful but she provides him no other information
(this requires one bit only). In case of the fully entangled state -- in 
quarter of the attempts she gets a singlet and succeed 
to teleport Carl's state (and Bob need not do anything).
In case Alice and Bob use any other pure entangled state the probability of
successful teleportation is less than $\frac{1}{4}$ but the fidelity 
of the successful events is still one.
The result of any local hidden variables theorem is that Alice can
never succeed transmitting the state to Bob with fidelity $1$.
This can be used to analyze non-local properties of mixed states 
(such as the Werner states) as well.
However, this is beyond the scopes of this work.

\section*{Acknowledgement}

I would like to thank Asher Peres for helpful discussions.

\end{document}